# Post-Acceleration Study for Neutrino Super-beam at CSNS

WU Yang[1,2], TANG Jing-Yu[1*]
[1]Institute of High Energy Physics, CAS, Beijing 100049, China
[2]College Nuclear Science and Technology, University of Science and Technology of China, Hefei, China

Abstract: A post-acceleration system based on the accelerators at CSNS (China Spallation Neutron Source) is proposed to build a super-beam facility for neutrino physics. Two post-acceleration schemes, one using superconducting dipole magnets in the main ring and the other using room-temperature magnets have been studied, both to achieve the final proton energy of 128 GeV and the beam power of 4 MW by taking 10% of the CSNS beam from the neutron source. The main design features and the comparison for the two schemes are presented. The CSNS super-beam facility will be very competitive in long-baseline neutrino physics studies, compared with other super-beam facilities proposed in the world.

## 1. Introduction

There are increasing interests worldwide in building new neutrino experimental facilities, since neutrino-related physics is key in the modern particle physics and astrology. There are different neutrino sources for experimental neutrino studies: cosmic neutrinos, reactor-based neutrinos and accelerator-based neutrinos. Although very expensive in building high beam power hadron accelerators, it is still considered very attractive using accelerator-based neutrinos. There are three major steps towards full exploration of accelerator-based neutrinos: present neutrino facilities, such as T2K [1] at J-PARC, CNGS [2] at CERN, NOvA [3] at FNAL, the next super-beams such CNGS+ [4], Project-X [5] and recently proposed LAGUNA-LBNO [6-7] etc, and the long-term plans such as Beta Beam [8], Neutrino Factory or Muon Collider [9]. This paper presents the post-acceleration studies based on the CSNS (China Spallation Neutron Source) accelerators which are under construction [10], for the purpose of building a super-beam facility.

A super-beam is a conventional high intensity neutrino beam produced by pion and kaon decays. In order to produce super-beam, one needs a high power proton accelerator complex which can provide MW-class proton beam whose energy ranges from several GeV to several hundred GeV. When proton beams are fired into target, kaons and pions are produced. After that, high intensity neutrinos are released in K and π decays. The two main features of a super-beam facility are: neutrinos are produced by high power accelerators (>1MW) and they are detected by large-scale detectors.

CSNS is a multi-disciplinary research facility, mainly for matter structure studies using neutron scattering techniques. As it is shown in Figure 1, the CSNS accelerator complex consists of an H$^-$ Linac of 250 MeV in the final phase, a proton rapid cycling synchrotron (RCS) of 1.6 GeV and beam transport lines, with the ultimate beam power of 500 kW. Table 1 shows the main parameters of the CSNS accelerators in different phases. The post-acceleration scheme studies for a super-beam facility are based on the CSNS-II' accelerators.

---

* Corresponding author: tangjy@ihep.ac.cn

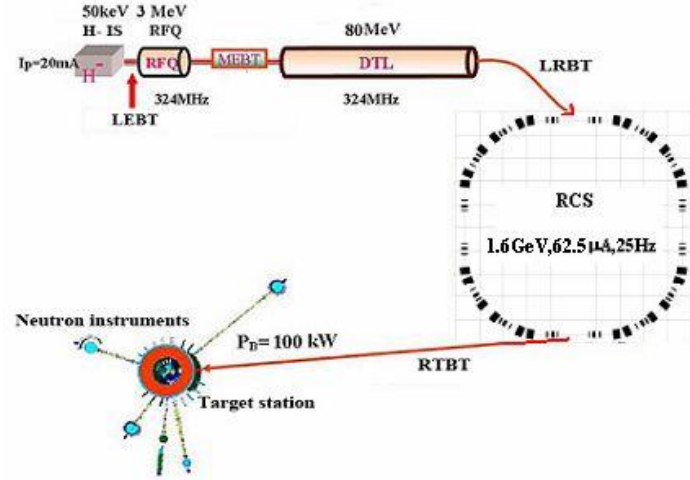

Figure 1: Schematic layout of CSNS

Table 1 Main parameters of the CSNS accelerator complex

|  | CSNS-I | CSNS-II | CSNS-II′ |
|---|---|---|---|
| Beam power [kW] | 100 | 200 | 500 |
| Proton energy [GeV] | 1.6 | 1.6 | 1.6 |
| Average beam current [μA] | 62.5 | 125 | 315 |
| Repetition rate [Hz] | 25 | 25 | 25 |
| Protons per pulse [$10^{13}$] | 1.56 | 3.12 | 7.8 |
| RCS circumference [m] | 227.92 | 227.92 | 227.92 |
| RCS lattice structure | S=4, all triplets | S=4, all triplets | S=4, all triplets |
| RF harmonics | 2 | 2 | 2 |
| Linac energy [MeV] | 80 | 130 | 250 |
| Linac peak current [mA] | 20 | 30 | 40 |
| Linac RF frequency [MHz] | 324 | 324 | 324 |
| Linac duty factor [%] | 0.5 | 1.0 | 1.5 |

## 2 Accelerator scheme design for CSNS super-beam

### 2.1 Design goal and overall design considerations

The objective of this study is to design a post-acceleration system based on the accelerator complex of CSNS (Phase II' as shown in Table 1), which will be the proton source of a super-beam facility for neutrino oscillation studies. The post-acceleration complex designed so far includes a booster, an accumulator ring (AR) and a main ring (MR). Proton beams will be accelerated to the energy of 128 GeV and the beam power of 4 MW. Two different schemes have been studied, with one using superconducting magnets in the main ring and the other using room temperature magnets in the main ring. For both schemes, only 10% of the CSNS beam is used for the post acceleration, in order to minimize the influence to the CSNS applications. In the following, the overall design considerations are presented.

**1) Final proton energy**
The power factor $F$ is defined as the product of proton energy and beam intensity:
$$F \triangleq E_p \times N_{pot} \tag{1}$$
In order to observe the $\nu_\mu \to \nu_e$ oscillation, $F$ is supposed to reach the order of $5 \times 10^{23}$ GeV.pot/year, with "pot" denoting protons on target. It has been found that neutrino event rate in the relevant region scales approximately linearly with the proton energy [11]. We know that different proton energies will produce different neutrino energy spectra, which directly influence the probability of $\nu_\mu \to \nu_e$ oscillation event. $F$ is an approximate measurement of proton economy and lower power factor means better proton economy.

In fact, for different baseline designs different proton energies are required, the longer baseline the higher proton energy. For example, proton energy of 20 GeV appears to be the most economical choice for the baseline of 730 km [12]. However, in the case of CSNS super-beam, to minimize the influence to the CSNS applications in matter structure studies, the beam intensity provided to the super-beam facility is limited to $1.95 \times 10^{14}$ pps, which is 10% of the whole protons from the RCS. We need higher proton energy to reach the goal of $5 \times 10^{23}$ GeV.pot/year. Then the maximum proton energy of the post-acceleration system is designed to be 128 GeV. With 10% of CSNS proton beam, the final beam power will reach 4 MW. If lower energy proton beam is used for neutrino studies, the beam power is also lower.

**2) Two-step acceleration**

As presented above, the final proton energy of the CSNS super-beam is chosen to be 128 GeV, while the output energy of RCS is only 1.6 GeV. This means that the post-acceleration should booster the beam energy by 80 times. This can be achieved only by two-step accelerations instead of one-step, with the first step called Booster and the second step called Main Ring. A more detailed study suggests that the output energy of 20 GeV from the Booster is almost an optimized solution, with the energy gains of 12.5 times in the Booster and 6.4 times in the Main Ring. The extraction energy from the Main Ring can be easily adjusted according to the needs of neutrino studies.

**3) Using an accumulator ring**

The study also shows that an accumulator ring between the Booster and the Main Ring is very helpful to reduce the cost and the technical difficulties of the Main Ring. In each cycle, the first beam batches from the Booster will be injected into the AR to store for a while, when the Booster accelerates the next beam batches. When the last beam batch reaches the extraction energy, the stored beam in the AR will be extracted and injected into the Main Ring, and it is followed by injecting the last beam batch directly from the Booster. The use of AR can lower the repetition rate or reduce the waiting time of the Main Ring, which is extremely important. Because there is no acceleration in the AR, and the Booster and AR can share a tunnel with a stacking layout or one above the other, the cost of adding AR is fully acceptable. There are two important benefits from lowering the ramping rate of the dipole magnets in the Main Ring. The first is about the reduction in the RF cavities used, and this can be seen from the following formula:

$$V\sin\varphi_s = \rho L (dB/dt) \tag{2}$$

where $V$ is the total RF voltage, $\varphi_s$ the synchronous phase, $\rho$ the curvature radius in dipole magnets, $L$ the ring circumference, $dB/dt$ is the dipole ramping rate.

The second is about the ramping rate itself when superconducting dipole magnets are used in the Main Ring. The state-of-art ramping rate is about 2 T/s for high field superconducting magnets, probably 4 T/s for magnetic field lower than 2 T [13].

**4) Main Ring with room-temperature magnets or superconducting magnets**

Superconducting dipole magnets (SC-magnets) have many advantages over ordinary room-temperature dipole magnets (RT-magnets). They can provide higher magnetic field than room-temperature magnets that are limited to fields around 1.6 T. With higher magnetic field, space can be saved so that a more compact ring can be designed and many RF cavities can be saved, as Eq. (2) suggests. Meantime, they consume much less electricity than room temperature magnets. However, SC-magnets have some drawbacks as well. Compared with RT-magnets, the maximum ramping rate for SC-magnets is much lower; SC-magnets also require more complicated fabrication techniques. Therefore, it is not so obvious if we should use SC-magnets or RT-magnets in the Main Ring. In the following, two different schemes with one using RT-magnets (called RT-scheme or RT-MR) and the other using SC-magnets (called SC-scheme or SC-MR) will be presented.
Also due to the requirement on fast ramping, SC-magnets are not considered in designing the Booster.

## 2.2 Post-acceleration scheme with RT-magnets in the Main Ring

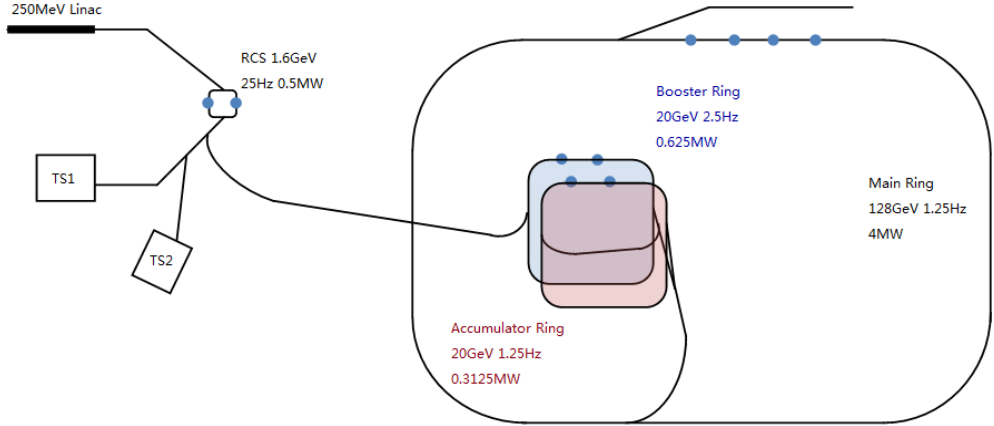

Figure 2: CSNS post-acceleration scheme with the Main Ring using RT-magnets. Small filled circles in the rings are for numbers of beam bunches.

Figure 2 shows the layout of the CSNS post-acceleration with the Main Ring using RT-magnets. The post-acceleration system is made up of three major parts: the Booster, the Accumulator Ring and the Main Ring. Each beam pulse extracted from RCS contains two bunches. As mentioned above, the post-acceleration system uses 10% of the total proton beam, in other words, the Booster receives one pulse in every ten pulses extracted from RCS. Protons in bunches will be accelerated to 20 GeV in the Booster. Then, they will be injected into the Accumulator Ring. The beam in the Accumulator Ring will be stored for 400 ms, during which the Booster will accelerate another two bunches to 20 GeV. Then after, the two beams in the Booster and the Accumulator will be extracted with a slight time difference and injected into the Main Ring. Eventually, the proton beam will be accelerated to 128 GeV in the Main Ring. The main parameters of the three accelerators are summarized in Table 2.

Table 3: Main parameters of the Booster, AR and MR for RT-MR scheme

|  | Booster | AR | MR |
|---|---|---|---|
| Circumference [m] | 1025.64 | 1025.64 | 4558.4 |
| Protons per pulse [$10^{13}$] | 7.8 | 7.8 | 15.6 |
| Input energy [GeV] | 1.6 | 20 | 20 |
| Output energy [GeV] | 20 | 20 | 128 |
| Beam power [MW] | 0.625 | 0.3125 | 4 |
| Magnetic rigidity [Tm] | 7.867- 73.111 | 73.111 | 73.111-430.1 |
| Repetition rate [Hz] | 2.5 | 1.25 | 1.25 |
| RF scanning frequency range [MHz] | 2.4444-2.6280 | 2.6280 | 2.6280-2.6306 |
| Curvature radius [m] | 54.41 | 54.41 | 290.20 |
| Circumference over bending segment | 3 | 3 | 2.5 |
| Dipoles field (input-output) [T] | 0.1446-1.3437 | 1.3437 | 0.2519-1.4820 |

| Dipole ramping rate during rising period [T/s][1] | 3.075-4.919 | - | 2.050 |
| --- | --- | --- | --- |
| RF harmonics | 9 | 9 | 40 |
| RF voltage [kV][1] | 343.16 | 44 | 3131.4 |
| Number of ferrite cavities (or MA cavities)[2] | 16 (8) | 2 (1) | 143 (70) |
| Transverse acceptance [πmm.mrad][3] | 180 | 20 | 20 |

[1] The dipoles field rises in the first 3/4 cycle and falls in the last 1/4 cycle. During the rising period, synchronous phase is set to 30° in the first half and 53° in the second half for the Booster and 60° for the Main Ring.

[2] Ferrite cavities: with RF voltage 22 kV per cavity; Magnetic alloy (MA) cavities: with RF voltage 45 kV per cavity.

[3] We assume that the transverse acceptance is 1.5 times the emittance. The full emittance of proton beam from the RCS is assumed to be 250 πmm.mrad. After passing through the beam line, proton beams are collimated and the emittance decreases to 120 πmm.mrad.

Figure 3 shows how dipole fields change over time and the beam extractions in the four accelerators. For comparison, two scenarios show what would happen in the MR if there were no accumulator ring, where Label 1 (in dashed line) denotes for MR with the repetition rate of 2.5 Hz and Label 2 (in dotted line) for MR with the repetition rate of 1.25 Hz.

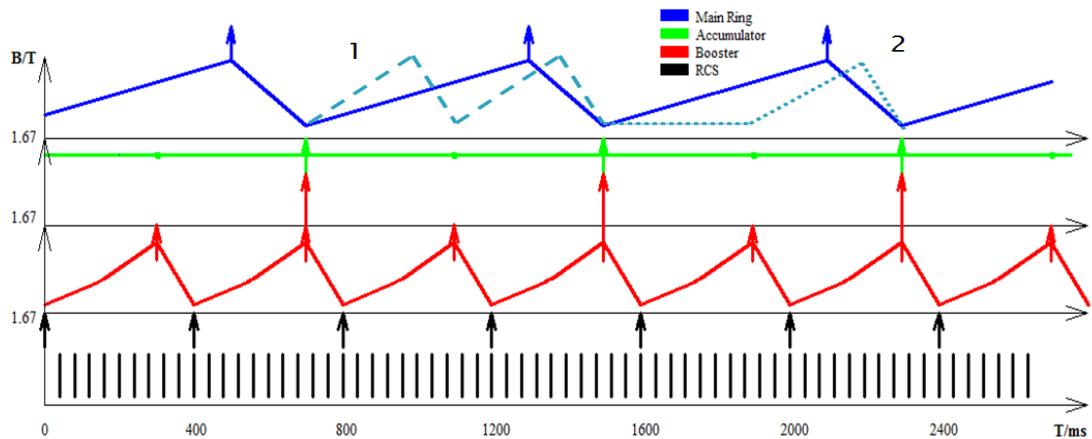

Figure 3: Field patterns and beam extractions in the four accelerators for the post-acceleration scheme with RT-MR

## 2.3 Post-acceleration scheme with SC-magnets in the Main Ring

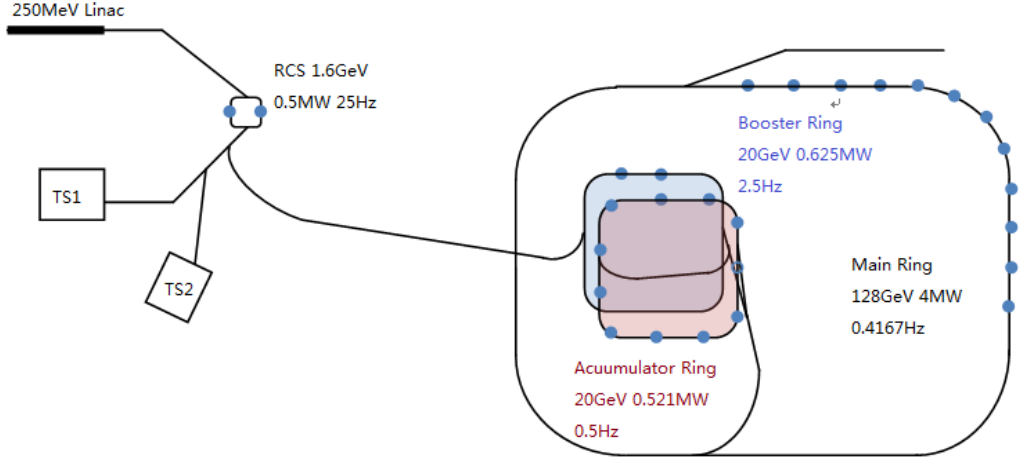

Figure 4: Post-acceleration scheme with the Main Ring using SC-magnets. Small filled circles in the rings are for numbers of beam bunches, and empty circles in AR denote the empty bucket.

Figure 4 shows the layout of the post-acceleration scheme with the Main Ring using SC-magnets for the CSNS super-beam. Similar to the scheme with RT-magnets, the post-acceleration system is also made up of three major parts: the Booster, the Accumulator Ring and the Main Ring. The main difference is on the repetition rate of MR. As mentioned above, lower repetition rate in MR is required to adopt SC-magnets and reduce the number of RF cavities. The circumference of AR is enlarged to receive more bunches from the Booster, which is key in reducing the repetition rate of MR. The circumference of the Booster is also enlarged to share the same tunnel with AR. The circumference of MR is shortened with the use of SC-magnets. The main parameters of the three accelerators are summarized in Table 4. Figure 5 shows Field patterns and beam extractions in the four accelerators. More detailed comparison between the two schemes is given in the next section.

Table 4: Main parameters of the Booster, AR and MR for the SC-MR scheme

|  | Booster | AR | MR |
|---|---|---|---|
| Circumference [m] | 1253.56 | 1253.56 | 3418.8 |
| Protons per pulse [$10^{13}$] | 7.8 | 39 | 46.8 |
| Input energy [GeV] | 1.6 | 20 | 20 |
| Output energy [GeV] | 20 | 20 | 128 |
| Beam power [MW] | 0.625 | 0.5208 | 4 |
| Magnetic rigidity [Tm] | 7.867- 73.111 | 73.111 | 73.111-430.1 |
| Repetition rate [Hz] | 2.5 | 0.5 | 0.4167 |
| RF scanning frequency range [MHz] | 2.4444-2.6280 | 2.6280 | 2.6280-2.6306 |
| Curvature radius [m] | 66.5 | 66.5 | 215.04 |
| Circumference over bending segment | 3 | 3 | 2.53 |
| Dipoles field (input-output) [T] | 0.1183-1.0994 | 1.0994 | 0.3400-2.0000[5] |
| Dipole ramping rate during rising period [T/s][4] | 2.5156-4.0250 | - | ＞1.3833 |

| RF harmonics | 11 | 11 | 30 |
|---|---|---|---|
| RF Voltage [kV][1] | 419.38 | 44 | 1174.3 |
| Number of ferrite cavities (or MA cavities)[2] | 20 (10) | 2 (1) | 54 (27) |
| Transverse acceptance [πmm.mrad][3] | 180 | 20 | 20 |

[4]In order to reduce the ramping rate to the limitation of SC-dipole magnet, the raising period and falling period are both set as 1200 ms.
[5]The highest dipole field is 2 T in SIS100 FAIR (still in R&D period) which is the highest SC-dipole field around the world.

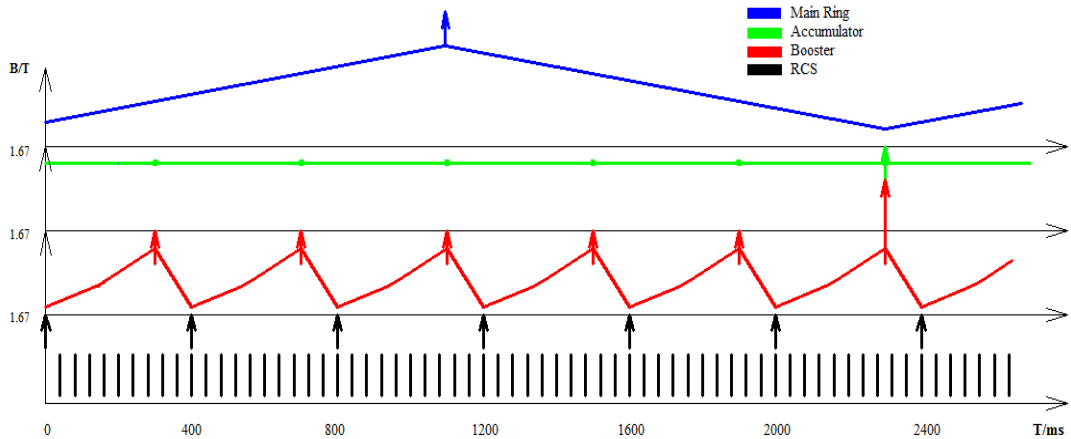

Figure 5: Field patterns and beam extractions in the four accelerators for the post-acceleration scheme with SC-MR

## 2.4 Comparison between the two schemes

With the same design goal, two different schemes presented above have their own advantages and disadvantages. Here some comparison is given to find out which scheme is more suitable for the CSNS post acceleration.

In both schemes the ramping-up time is designed to be longer than the ramping-down time in both the Booster and the Main Ring, and this helps reduce the number of RF cavities significantly. However, with the future development of superconducting magnets, the applicable field ramping rate can be higher, thus there will be probably more advantages on the SC-magnet scheme.

As the cost of the Accumulator Ring weighs quite low, Tables 5 and 6 show only the differences on the Booster and the Main Ring for the two post-acceleration schemes.

Table 5: Comparison of the Booster between the RT-MR scheme and the SC-MR scheme

| Booster parameters | With RT-MR | With SC-MR |
|---|---|---|
| Circumference [m] | 1025.64 | 1253.56 |
| Dipole field [T] | 0.1446-1.3437 | 0.1183-1.0994 |
| Dipole ramping rate during rising period [T/s] | 3.0746-4.9194 | 2.5156-4.0250 |
| Curvature radius [m] | 54.41 | 66.50 |
| RF Voltage [kV] | 343.16 | 419.38 |
| Number of ferrite cavities (or MA cavities) | 16 (8) | 20 (10) |

| RF harmonics | 9 | 11 |

Table 6: Comparison of the Main Ring between the RT-MR scheme and the SC-MR scheme

| Main Ring parameters | With RT MR | With SC MR |
|---|---|---|
| Circumference [m] | 4558.40 | 3418.8 |
| Repetition rate [Hz] | 1.25 | 0.4166 |
| Dipole field [T] | 0.2519-1.4820 | 0.3400-2.0000 |
| Dipole ramping rate during rising period [T/s] | 3.0753 | 1.3833 |
| Curvature Radius [m] | 290.20 | 215.04 |
| RF harmonics | 40 | 30 |
| RF voltage [kV] | 3131.4 | 1174.3 |
| Number of ferrite cavities (or MA cavities) | 143 (71) | 54 (27) |

From Tables 5 and 6, we can draw the following conclusions:
1) About the Booster: no scheme shows evident advantages.
The circumference of the Booster in the SC-MR scheme is 1.2 times as long as the one in the RT-MR scheme. This is mainly due to that the Accumulator Ring is longer in order to accumulate more pulses, and the Booster will share the same tunnel with the Accumulator Ring.
The Booster in the SC-MR scheme requires four more RF cavities, and this is also due to the longer circumference.
2) About the Main Ring: the SC-MR scheme has obvious advantages.
Since superconductive magnets can generate higher magnetic field in dipoles, the SC Main Ring has shorter circumference or less focusing periods. Thus, the SC Main Ring has advantage in reducing the costs of tunnel, quadrupoles, diagnostics, and vacuum etc. More important, the number of RF cavities is reduced largely by both slower cycling and shorter circumference, resulting in only one third compared with the RT-MR.
In addition, superconductive magnets consume much less power than room-temperature magnets so that the operation cost can be reduced, with the additional investment in the cryogenic system.
From the above analysis, the SC-MR scheme is favored.

## 3 Comparison with other super-beam facilities

In order to show the competitive feature of the CSNS super-beam facility, Table 7 shows the comparison among the six major super-beam projects in the world, and all of them are in the design phase.

Table 7: Main parameters of major super-beam projects

| | Proton beam energy (GeV) | Protons per pulse | Repetition period (s) | Beam power (MW) |
|---|---|---|---|---|
| GNGS+, CERN | 400 | 4.8-14 $\times 10^{13}$ | 6 | 0.3-1.2 |
| Project-X, FNAL | 120 | 9.5-15 $\times 10^{13}$ | 1.5 | 1.1-2 |
| T2K+, J-PARC | 50 | 33 $\times 10^{13}$ | 3.64-1.6 | 0.6-1.5 |
| BNL Super-beam | 28 | 9-25 $\times 10^{13}$ | 0.4-0.1 | 1-4 |
| FREJUS, CERN | 3.5 | 14.3 $\times 10^{13}$ | 0.02 | 4 |
| LAGUNA, CERN | 50 | $25 \times 10^{13}$ | 1 | 2 |
| CSNS Super-beam | 128 | $46.8 \times 10^{13}$ | 2.4 | 4 |

From Table 7, we can see that for all the facilities the beam power is about several MW, while the proton beam energy ranges from several GeV to several hundreds of GeV.

With lower proton energy such as in the case of FREJUS, the proton beam accelerated by a superconducting linac - SPL has high intensity and high repetition rate. This makes it fit well to short baseline neutrino studies with relatively soft energy spectra. Encouraged by the successful experiments in CNGS and T2K, both facilities are proposed to be upgraded to become more competitive in the future.

By adding an 8-GeV superconducting linac to the existing Main Injector of 120 GeV, Project-X is a very competitive super-beam facility in the USA, very suitable for long baseline neutrino oscillation experiments.

Similar to Project-X, CSNS super-beam has comparable beam energy but with doubled beam power.This means CSNS super-beam can generate same amount of neutrino with Project-X with only half of the time, which make CSNS super-beam very competitive in long baseline neutrino oscillation study.

## 4. Conclusions

Two post-acceleration schemes for CSNS neutrino super-beam facility are studied, with the same design goal. Among them, the scheme with superconducting magnets in the Main Ring is favored due to lower construction and operation costs. Both schemes exploit only 10% of CSNS/RCS beams, so that the applications based on neutron scattering are almost not affected.

With the beam energy of 128 GeV and the beam power of 4 MW, the CSNS super-beam will be very competitive in long-baseline neutrino physics, compared with other proposals in the world.

## Acknowledgements

The authors want to thank Yifang Wang for initiating the study and other IHEP colleagues for discussions. Projects 11235012 and 10975150 supported by National Natural Science Foundation of China"